\documentclass[12pt]{iopart}
\usepackage{iopams}
\usepackage{graphicx,multirow}

\begin{document}

\title{Local structural distortions and their role in superconductivity in SmFeAsO$_{1-x}$F$_x$ superconductors}

\author{Kapil Ingle$^1$, K.R. Priolkar$^1$, Anand Pal$^2$, V. P. S. Awana$^3$ , S. Emura$^4$}

\address{$^1$ Department of Physics, Goa University, Goa 403206, India}
\address{$^2$ Department of Physics, Indian Institute of Science, Bangalore 560012, India}
\address{$^3$ National Physical Laboratory, Dr. K.S. Krishnan Marg, New Delhi 110012, India}
\address{$^4$ Institute of Scientific and Industrial Research, Osaka University, Osaka, Japan}
\ead{krp@unigoa.ac.in}

\begin{abstract}
EXAFS studies at the As K edge as a function of temperature have been carried out in SmFeAsO$_{1-x}$F$_x$ ($x$ = 0 and 0.2) compounds to understand the role of local structural distortions superconductivity observed in F doped compound. A significant correlation between thermal variation of local structural parameters like anion height and superconducting onset is found in the fluorinated compound. Such a variation in anion height is absent in non-superconducting compound. Increase in Fe-As bond distance just below superconducting onset temperature indicates a similarity in distortions observed in high T$_C$ cuprates and these Fe based superconductors.
\end{abstract}

\maketitle

\section{Introduction}
Discovery of superconductivity in Fe based systems has initiated a flurry of research activity due to (a) high superconducting transition temperature (T$_C \sim$ 50K); (b) similarity in local structure of Fe in all these iron based superconductors;  and (c) their similarity of structure and phase diagrams with those exhibited by cuprates and heavy Fermion superconductors \cite{scal}. Structurally, all Fe based superconductors consist of a stack of Fe - X networks (where X is either a chalcogen or pnictogen ion) separated with or without a spacer layer. Here Fe ions form a planar two dimensional square lattice and the pnictide or chalcogen ions occupy the centres of these square displaced alternatively above and below the plane of Fe atoms.  Again the stacking can be of various types like in REOFeAs (RE- rare earth), the REO layers act as a spacer layer \cite{kami,ren} the Ba ions play the role of spacers in BaFe$_2$As$_2$ type compounds \cite{sefat} while there are no spacer layers in FeSe type superconductors \cite{mizu,oka}.

Superconducting T$_C$ shows a considerable enhancement under the influence of hydrostatic or chemical pressure. For example, T$_C$ enhances to 43K when a pressure of 4 GPa is applied to LaFeAsO$_{1-x}$F$_x$ \cite{taka}. Similarly changing the rare earth ion from La to Sm also leads to an increase in T$_C$ due to pressure exerted by smaller Sm ion \cite{ren}. Superconductivity can also be induced due to doping of holes or electrons as in case of Ba$_{1-x}$K$_x$Fe$_2$As$_2$ (hole doping) \cite{sefat} or in BaFe$_{1.8}$Co$_{0.2}$As$_2$ (electron doping) \cite{chu} as well as by changing the chalcogen ion as in FeTe$_{1-x}$Se$_x$ \cite{liu}. All these results indicate a strong influence of local structure especially around Fe ion on superconducting interactions in these compounds. Such an influence of local structure has also been reported in high T$_c$ cuprates and is believed to be intimately connected with the mechanism of superconductivity in these materials \cite{bil,saini}.

Local structure around Fe atoms has been obtained in different families of iron based superconductors as a function of temperature, pressure and chemical substitution \cite{zhang,iade,joseph,granado}.  In substituted compounds, especially of the type FeSe$_{1-x}$Te$_x$, coexistence of phases (FeSe and FeTe) separated over nanoscopic dimensions have been noticed \cite{saini1}. Anion height ($h$) which is defined as the perpendicular distance between the plane of anions and the plane of Fe atoms has emerged as an important parameter that can be correlated with superconductivity in these compounds. Lower values of $h$ directly influence the hybridization between the Fe 3d bands and the anion p bands and thereby regulating the nature of Fermi surface \cite{subedi}. EXAFS studies have also revealed that application of hydrostatic pressure and/or modification of intercalated layers cause sizeable compression of Fe-X networks (reduction in $h$) \cite{joseph, granado}. Such modifications are known to influence T$_C$, for example in Ba$_{0.8}$K$_{0.2}$Fe$_2$As$_2$ \cite{rotter}. Fluctuations in local lattice order near the onset of superconductivity (T$_{C-onset}$) have been reported through a temperature dependent EXAFS study in LaFeAsO$_{1-x}$F$_x$ compounds \cite{zhang}. All these studies point towards a strong role of electron-lattice interactions in superconductivity.

However, it is still not clear if the local structural distortions have a role to play in superconductivity. For instance, it is known that anion height is much smaller in superconducting Fe based compounds as compared to their non-superconducting counterparts. But does this anion height change near the onset of superconductivity? Is there any correlation between local structural distortions in Fe based superconductors and other superconducting compounds like cuprates? In order to seek answers to these questions and to establish a correlation between local structural distortions and superconductivity in FeAs based superconductors, we have carefully studied As K edge EXAFS as a function of temperature in superconducting (x = 0.2) and non-superconducting (x = 0) SmFeAsO$_{1-x}$F$_x$ compounds.

\section{Experimental}
Bulk polycrystalline SmFeAsO and SmFeAsO$_{0.8}$F$_{0.2}$ compounds were synthesized by conventional solid state reaction route method via vacuum encapsulation technique.  The high purity ($\sim$99.9\%)  Sm, As, Fe, FeF$_3$ and Fe$_2$O$_3$ reactants were weighed according to the stoichiometric ratio, mixed and ground thoroughly using mortar and pestle. The weighing and grinding was carried out in a glove box of $<$ 1ppm level of Oxygen and humidity atmosphere. The mixed powder was pelletized in rectangular bar shape and then encapsulated in an evacuated (10$^{-3}$ Torr) quartz tube. These sealed quartz tubes embodying the said pellet were heat treated at 550$^\circ$C for 12 h, 850$^\circ$C for 12 h and then at 1150$^\circ$C for 33 h in continuum, then switched off the furnace and allowed to cool naturally. The sintered sample was obtained by breaking the quartz tube. The as sintered sample was black in color and brittle. For transport measurements, the obtained sample is again ground and sealed in quartz tube and finally heated for 12 hours at 1150$^\circ$C with a slow heating rate to obtain a good compact pellet. Finally the samples were furnace cooled slowly to room temperature\cite{awana}.

Phase identification and crystal structure investigation were carried out by using powder X-ray diffraction (Rigaku-XRD) with Cu-K$_\alpha$ radiation.  The lattice parameters are calculated with the help of Rietveld refinement using the FULLPROF SUITE program. The resistivity measurements were performed by a conventional four-point-probe method on a Quantum Design Physical Property Measurement System (PPMS-140kOe). The DC-magnetization susceptibility measurement was performed on superconducting sample using DC magnetometry system (ACMS-Model-P500) on same QD-PPMS.

EXAFS measurements at the As K edge were performed in transmission mode at BL09C beamline at Photon Factory, Japan. EXAFS was scanned from -200 eV to 1200 eV with respect to As K edge energy (11600 eV).  Both incident (I$_0$) and transmitted (I) intensities were measured simultaneously using ionization chamber filled with appropriate gases. The absorbers were prepared by sprinkling finely ground powder on scotch tape and stacking several such layers to optimize the thickness so that the edge jump ($\Delta\mu$) was restricted to $\le$ 1.

EXAFS data analysis in the k range of 2 to 14 \AA$^{-1}$  and in the R range of 1 to 3.5 \AA~ was performed using Demeter program \cite{ravel}. Theoretical amplitude and phase shift functions for different correlations were calculated using FEFF 6.01 \cite{rehr} using the crystal structure data obtained from room temperature X-ray diffraction.

\section{Results and Discussion}
\begin{figure}
\centering
\includegraphics[width=\columnwidth]{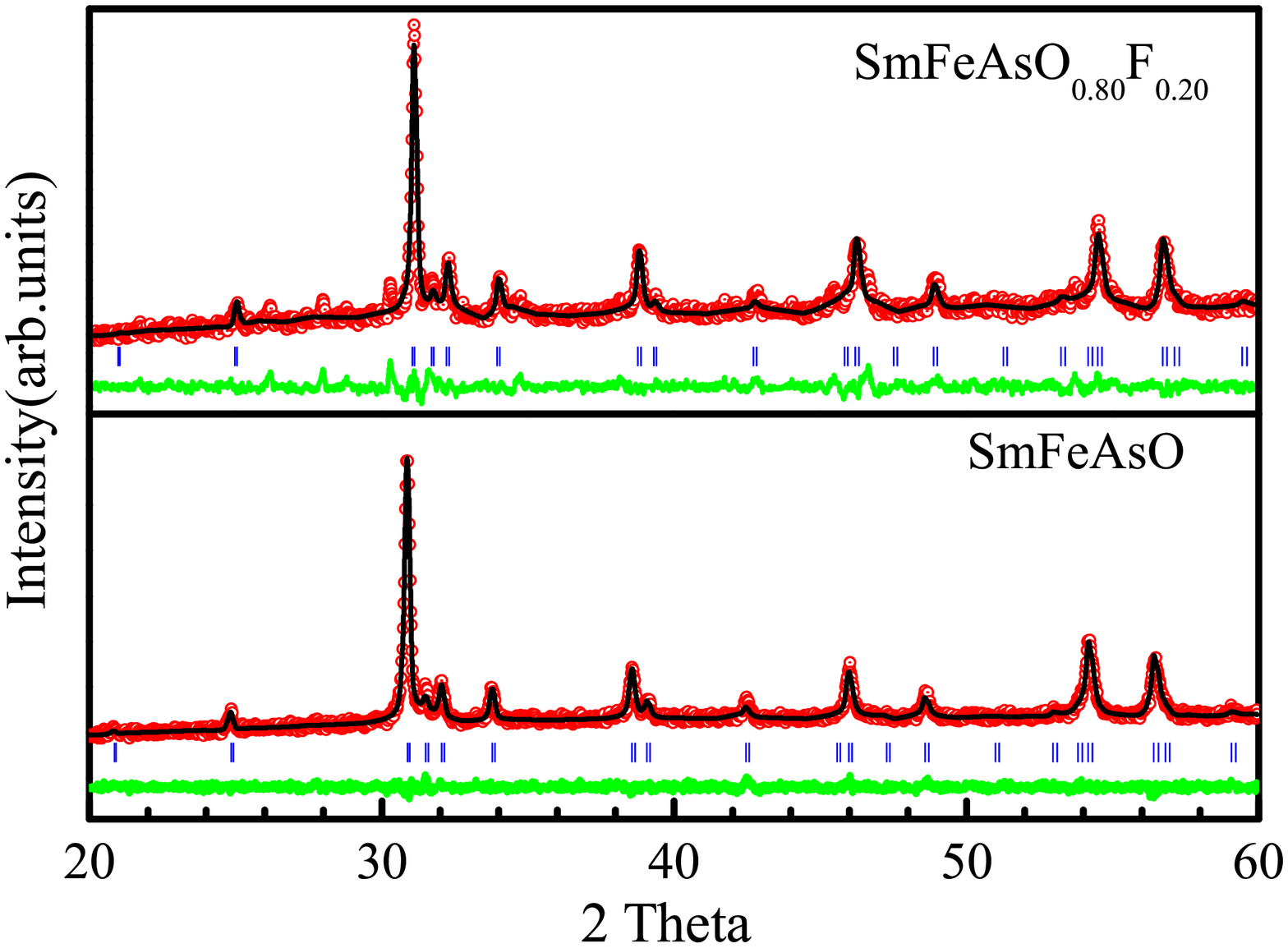}
\caption{\label{fig1}(a): Rietveld fitted room temperature X-ray diffraction patterns of SmFeAsO and SmFeAsO0.8F0.2 samples. All the permitted diffraction planes are marked with blue vertical lines between the observed, Yobs (red open circle)/fitted, Ycalc (black line) patterns and their difference Yobs- Ycalc (green line) in the bottom.}
\end{figure}

Observed and Rietveld fitted X-ray diffraction (XRD) patterns for polycrystalline SmFeAsO and SmFeAsO$_{0.8}$F$_{0.2}$ samples are shown in Fig. \ref{fig1}. The Rietveld analysis of the room temperature X-ray diffraction pattern confirmed that all observed reflections could be satisfactorily indexed on the basis of tetragonal crystal structure (space group: $P4/nmm$), ensuring  the phase purity of the studied sample. The Rietveld fitted pattern shows that the studied samples are nearly single phase except the additional weak peaks (marked with * in the XRD pattern), which were assigned to the rare earth oxide and fluoride impurity phases in the F-doped sample. The Rietveld refined structural and fitting goodness parameter along with Wyckoff position of atoms are shown in Table \ref{feas-tab1}. After F substitution very small reduction is observed in $a$ while $c$ reduces significantly. The contraction of $c$ in SmFeAsO$_{0.8}$F$_{0.2}$ sample indicates a successful substitution of F$^{-1}$ (R$_{\rm F}$ = 1.33 \AA) at O$^{2-}$ (R$_{\rm O}$ = 1.40\AA) sites. These evaluated values of lattice parameters of the synthesized samples are close to those reported earlier \cite{vgh,mart}. It is also clear from the Table \ref{feas-tab1} that the position of As atom is unaffected in both the sample, whereas the Sm atom shifted along the $c$-axis, because F$^{-1}$ substitution (at O$^{2-}$ site) deceases the negative charge at O site thereby increasing the polarization and coulomb interaction between the layers. The contraction of the Sm-As bond length (which is bridging the layers) coupled with a slight increase of the Sm-O bond length confirms the electronic polarization in F-doped sample (as shown in Table \ref{feas-tab2}). As a result the Sm-O and Fe-As tetrahedral layers come closer and the cell parameter $c$ decreases. \cite{omura}. The O-Sm-O and the As-Fe-As tetrahedral angle also decrease to maintain the perfect tetrahedral, consequently $a$ only changes slightly. The entire bond length and bond angle are calculated using bond valance sum method \cite{brese}. The observed structural results are fairly consistent with the reports on other rare earth substitutions, and indicate the covalent character of the intra-layer chemical bonding due to the smaller covalent radius of fluorine than oxygen \cite{mart,omura}.

\begin{table}
\caption{\label{feas-tab1}Reitveld refined Structural data of SmFeAsO and SmFeAsO$_{0.8}$F$_{0.2}$ samples with space group P4/nmm}

\begin{indented}
\lineup
\item[]\begin{tabular}{@{}*{9}{c}}
\br
Sample & \multicolumn{5}{c}{SmFeAsO} & \multicolumn{3}{c}{SmFeAsO$_{0.8}$F$_{0.2}$}\\
\mr
\multirow{3}{*}{Lattice parameters} & $a$\AA & \multicolumn{4}{c}{3.937(2)} & \multicolumn{3}{c}{3.926(2)}\\
& $c$\AA & \multicolumn{4}{c}{8.491(8)} & \multicolumn{3}{c}{8.460(4)}\\
& V \AA$^3$ & \multicolumn{4}{c}{131.637} & \multicolumn{3}{c}{130.417}\\
\mr
\multirow{5}{*}{Wyckoff Positions} &  & & x & y & z & x & y & z \\
\mr
& Sm & 2c & 0.25 & 0.25 & 0.1371(1) & 0.25 & 0.25 & 0.142(6)\\
& Fe & 2b & 0.75 & 0.25 & 0.5 & 0.75 & 0.25 & 0.5\\
& As & 2c & 0.25 & 0.25 & 0.661(5) & 0.25 & 0.25 & 0.660(2)\\
& O/F & 2a & 0.75 & 0.25 & 0 & 0.75 & 0.25 & 0\\
\mr
Bragg R-factor & \multicolumn{5}{c}{2.22}& \multicolumn{3}{c}{4.61}\\
$\chi^2$ &\multicolumn{5}{c}{2.02}& \multicolumn{3}{c}{3.07}\\
\br
\end{tabular}
\end{indented}
\end{table}

\begin{table}
\caption{\label{feas-tab2}Selected bond lengths and bond angles in SmFeAsO and SmFeAsO$_{0.8}$F$_{0.2}$.}
\begin{indented}
\lineup
\item[]\begin{tabular}{@{}*{3}{c}}
\br
Bond distance/Angle & SmFeAsO & SmFeAsO$_{0.8}$F$_{0.2}$\\
\mr
Sm-As & 3.2669 & 3.3287\\
Sm-O & 2.2871 & 2.3042\\
Fe-As & 2.3997 & 2.3958 \\
O-Sm-O & 118.80 & 116.85\\
As-Fe-As & 109.13 & 108.84\\
\br
\end{tabular}
\end{indented}
\end{table}

\begin{figure}
\centering
\includegraphics[width=\columnwidth]{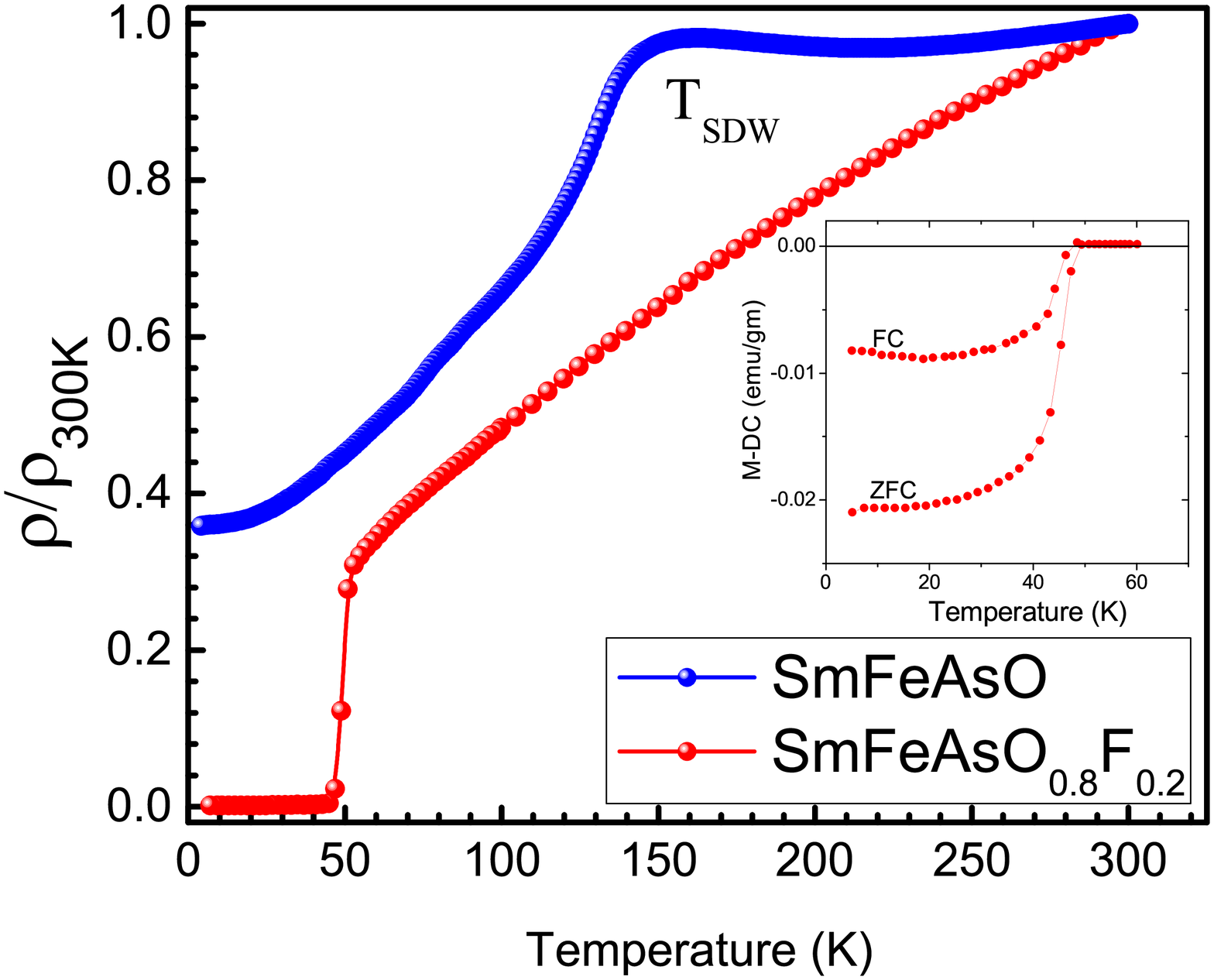}
\caption{\label{fig2}Resistivity behavior with temperature variation $\rho$(T) of  SmFeAsO and SmFeAsO$_{0.8}$F$_{0.2}$ samples.}
\end{figure}

The temperature dependence of normalized resistivity ($\rho$/$\rho_{300}$) for SmFeAsO and SmFeAsO$_{0.8}$F$_{0.2}$ samples are shown in Figure \ref{fig2}. The temperature dependent resistivity $\rho$(T) curve of undoped SmFeAsO sample shows an anomaly behaviour at around $\sim$ 150 K, it exhibits a semiconducting behaviour above the said temperature and suddenly resistivity becomes metallic in nature at lower temperature.  This peculiar behaviour is a coupled result of crystallographic phase transition from the tetragonal P4/nmm to the orthorhombic Cmma space group around T $\sim$150K, and the occurrence of static spin density wave (SDW) instability like magnetic ordering of the Fe spins at a slightly lower temperature of $\sim$140K \cite{chen,pal}. This feature disappears in the F-doped SmFeAsO$_{0.8}$F$_{0.2}$ sample. The resistivity in the F-doped sample is metallic in nature above the superconducting transition temperature T$_C$($\rho$ = 0) $\sim$49 K. The increasing metallic behaviour of F-doped sample indicates that the charge carrier density increases in the conduction plane. The superconducting transition width, $\Delta$T$_c$ (T$_{C-onset}$ - T$_C$($\rho$ = 0)), is found to be $\Delta$T$_c$ $\sim$ 2.9 K. The normal state resistivity shows linear dependence on temperature for the studied SmFeAsO$_{0.8}$F$_{0.2}$ sample \cite{zhi, meena}.

The Bulk superconductivity in the studied SmFeAsO$_{0.8}$F$_{0.2}$ sample is also confirmed by the DC magnetization measurements. The DC magnetic susceptibility versus temperature plot for the superconducting SmFeAsO$_{0.8}$F$_{0.2}$   sample in both zero-field-cooled (zfc) and field-cooled (fc) situations at H = 10 Oe are shown in the inset of fig. \ref{fig2}.  Superconductivity sets in below 50 K, as evidenced from the negative susceptibility in both ZFC and FC condition.  The slight difference in onset transition temperature obtained from resistivity $\rho$(T) and magnetization M(T) is due to the threshold for transport measurements.  The transport measurement is through percolation path and hence is lower in comparison to bulk diamagnetic response. The difference between ZFC and FC curves suggests that the material has a fairly large flux pinning force resulting in the trapping of magnetic flux under the field cooling condition.

\begin{figure}
\centering
\includegraphics[width=\columnwidth]{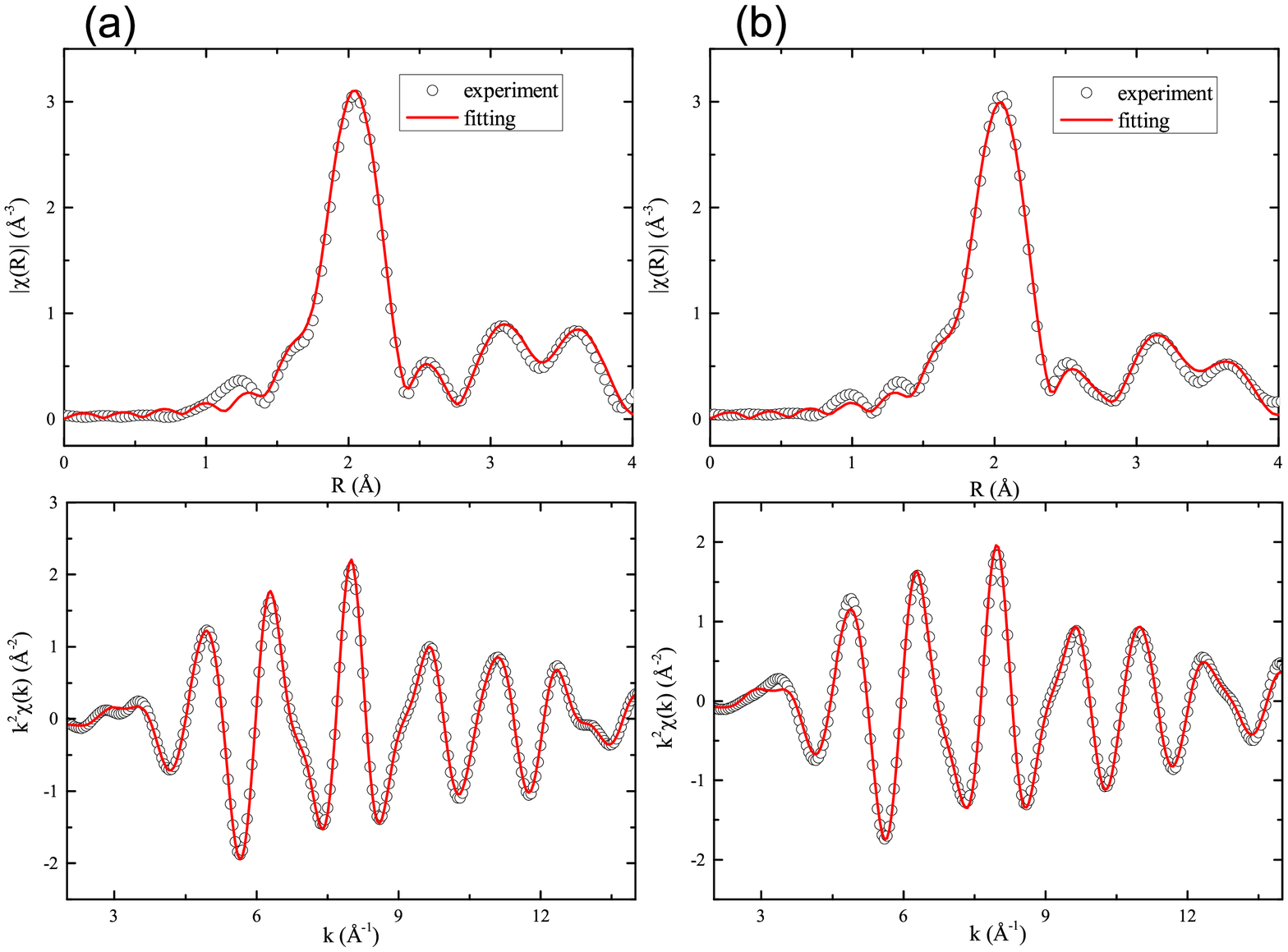}
\caption{\label{fig4} Magnitude of Fourier transform of $k^2$ weighted $\chi{k}$ recorded at room temperature in (a) SmFeAsO and (b) SmFeAsO$_{0.8}$F$_{0.2}$ samples.}
\end{figure}

Magnitude of the Fourier transform of EXAFS recorded at room temperature in the two compounds is presented in Figure \ref{fig4}. The data exhibits a prominent peak at about 2\AA~ corresponding to nearest neighbor As-Fe correlation. Two smaller humps corresponding to As-Sm, As-O and As-As correlations can be seen in the range 3 to 4 \AA. The data in the range 1 to 4 \AA~ was fitted to these three correlations. In order to reduce the number of fitting parameters and to extract all important anion height ($h$), these correlations were expressed in terms of a geometric relations based on lattice parameter $a$ and $h$. A good fitting was obtained in both R and k space as can be seen in Figure \ref{fig4}. The values of parameters obtained from fitting are presented in Table \ref{feas-tab3}.

\begin{table}
\caption{\label{feas-tab3}Selected bond lengths and bond angles in SmFeAsO and SmFeAsO$_{0.8}$F$_{0.2}$.}
\begin{indented}
\lineup
\item[]\begin{tabular}{@{}*{6}{c}}
\br
\multirow{2}{*}{Temperature} & \multirow{2}{*}{Bond $\times$ Coordination number} & \multicolumn{2}{c}{SmFeAsO} & \multicolumn{2}{c}{SmFeAsO$_{0.8}$F$_{0.2}$}\\
  & & R (\AA) & $\sigma^2$\AA$^2$ & R (\AA) & $\sigma^2$\AA$^2$\\
\mr
\multirow{5}{*}{300K} & As - Fe $\times$ 4 & 2.391(3) &0.006 (1) &2.394(1) &0.0055(1) \\
& As - Sm$\times$ 4 &3.243(7) &0.008 (2) &3.239(7) &0.011(1) \\
& As - O $\times$ 4 &3.30(7) &0.007 (3) & 3.33(4)& 0.020(8)\\
& As - As  $\times$ 4 & 3.89(1) &0.021(5) & 3.90(1)&0.014(1)\\
& As - As $\times$ 4 &3.93(1) & 0.021(5) & 3.93(1)& 0.014(1)\\
\mr
\multirow{5}{*}{100K} & As - Fe $\times$ 4  &2.392(3) &0.003 (1) &2.390(1) & 0.0032(1)\\
& As - Sm$\times$ 4 & 3.256(7)& 0.007(2)&3.234(6) & 0.007(1)\\
& As - O $\times$ 4 & 3.31(3)& 0.008(3)& 3.32(4)&0.020(8) \\
& As - As  $\times$ 4 & 3.88(1)& 0.011(1)& 3.89(1)& 0.008(1)\\
& As - As $\times$ 4 &  3.91(1)&0.011(1) & 3.92(1)& 0.008(1)\\
\mr
\multirow{5}{*}{60K} & As - Fe $\times$ 4 &2.392(1) &0.003 (1) &2.389(1) &0.0033(1) \\
& As - Sm$\times$ 4 & 3.261(7)& 0.007(1) &3.237(3) &0.006(1) \\
& As - O $\times$ 4 & 3.31(4)& 0.007(1) &3.32(3) & 0.013(5)\\
& As - As  $\times$ 4 & 3.88(1)&0.010(1) &3.90(1) & 0.007(1)\\
& As - As $\times$ 4 & 3.92(1)& 0.010(1) & 3.93(1)& 0.007(1) \\
\mr
\multirow{5}{*}{50K} & As - Fe $\times$ 4  & 2.392(1) & 0.003 (1) &2.390(1) &0.0033(1) \\
& As - Sm$\times$ 4 & 3.227(7)& 0.006(1) &3.237(4)&0.006(1) \\
& As - O $\times$ 4 & 3.24(4)& 0.004(3) & 3.32(4)& 0.013(6)\\
& As - As  $\times$ 4 & 3.88(1)&0.012(1) & 3.88(1)& 0.007(1)\\
& As - As $\times$ 4 & 3.92(1)&0.012(1) & 3.92(1)& 0.007(1)\\
\br
\end{tabular}
\end{indented}
\end{table}

\begin{figure}
\centering
\includegraphics[width=\columnwidth]{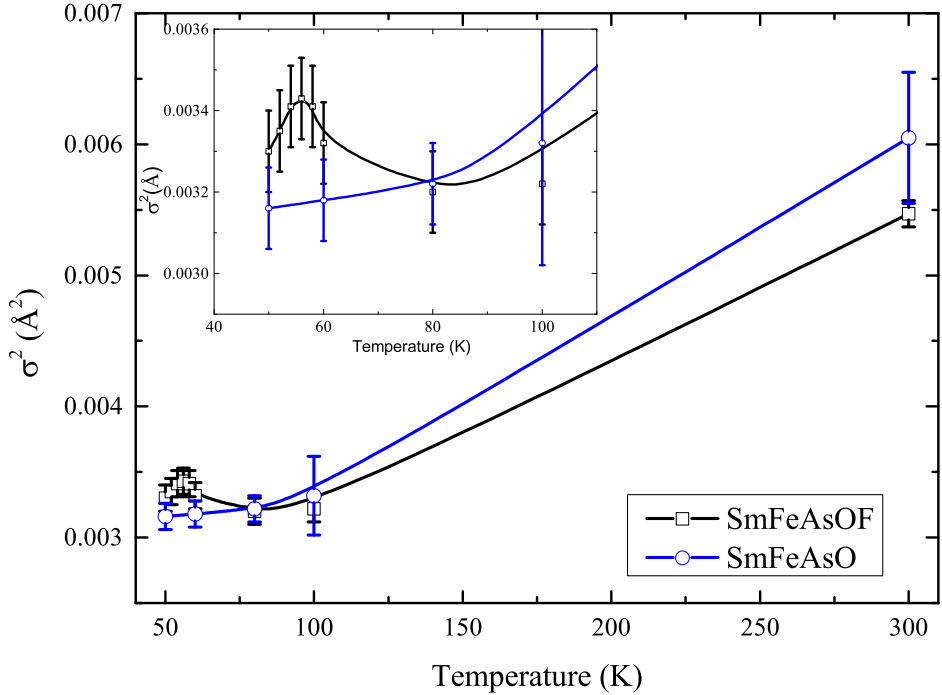}
\caption{\label{fig5} Variation of mean square disorder ($\sigma^2$) of As-Fe bond in SmFeAsO and SmFeAsO$_{0.8}$F$_{0.2}$ samples.}
\end{figure}

EXAFS studies on isostructural F doped and undoped LaFeAsO compounds have reported anomaly in mean square relative displacement (MSRD) or $\sigma^2$ of the nearest neighbor Fe - As around the T$_{C-onset}$ indicating presence of structural anomalies around the presence of superconducting transition. In Figure \ref{fig5}, temperature variation of $\sigma^2$ for nearest neighbor As-Fe is presented. The $\sigma^2$ value shows a decrease with temperature consistent with the Debye-type behavior. While a monotonic decrease seems to be present seen in case of non superconducting SmFeAsO, a clear upturn peaking at 56K is seen in the values of $\sigma^2$ in case of SmFeAsO$_{0.8}$F$_{0.2}$ compound. The temperature at which $\sigma^2$ exhibits a peak correlates well with the T$_{C-onset}$ temperature observed in resistivity. This is consistent with the observations reported in Ref. \cite{zhang}. In case of LaFeAsO sample the upturn was identified with a characteristic temperature T$^*$ and which was more than twice the T$_{C-onset}$. In case of SmFeAsO$_{0.8}$F$_{0.2}$, the upturn occurs at about 80K which is about 1.4 times the T$_{C-onset}$ temperature. In high T$_c$ superconductors, T$^*$ which corresponds to appearance of a pseudogap like phase, is about 1.5T$_c$.  Further, in the present case, it may be noted that apart from a change in the rare-earth ion which would affect the hybridization between Fe $3d$ and As $4p$ bands, there is also a difference in concentration of F. Fluorine doping in the compound studied here is about 20\% while in Ref. \cite{zhang} it was only 7\%. F doping is believed to introduce holes in the hybridized Fe-As band and hence our sample will be overdoped with holes as compared to LaFeAsO$_{0.93}$F$_{0.07}$. In the phase diagrams presented for other superconductor families like cuprates, the characteristic temperature T$^*$ decreases with increasing hole doping in CuO planes \cite{oya}.

The deviation of $\sigma^2$ from the expected Debye behavior in superconducting SmFeAsO$_{0.8}$F$_{0.2}$ suggests presence of dynamical distortions, appearing as split Fe-As bonds in EXAFS, as the temperature is decreased below T$^*$ to approach the superconducting transition. Such a splitting has been also proposed in F doped LaFeAsO material \cite{zhang}.

Anion height ($h$) is considered to be an important parameter for superconductivity in Fe based compounds as it directly related to the hybridization between Fe 3d and As 4p bands. Higher the hybridization, greater is the mobility of holes, thereby suppressing the antiferromagnetic interactions and ushering in superconductivity. Therefore, $h$ is smaller in case of compounds exhibiting superconductivity as compared to their non superconducting counterparts. A similar observation is made in F doped and undoped SmFeAsO samples. In the undoped sample the $h$ was found to be approximately 1.37\AA~ while in its superconducting variant $h \approx$ 1.364\AA.

\begin{figure}
\centering
\includegraphics[width=\columnwidth]{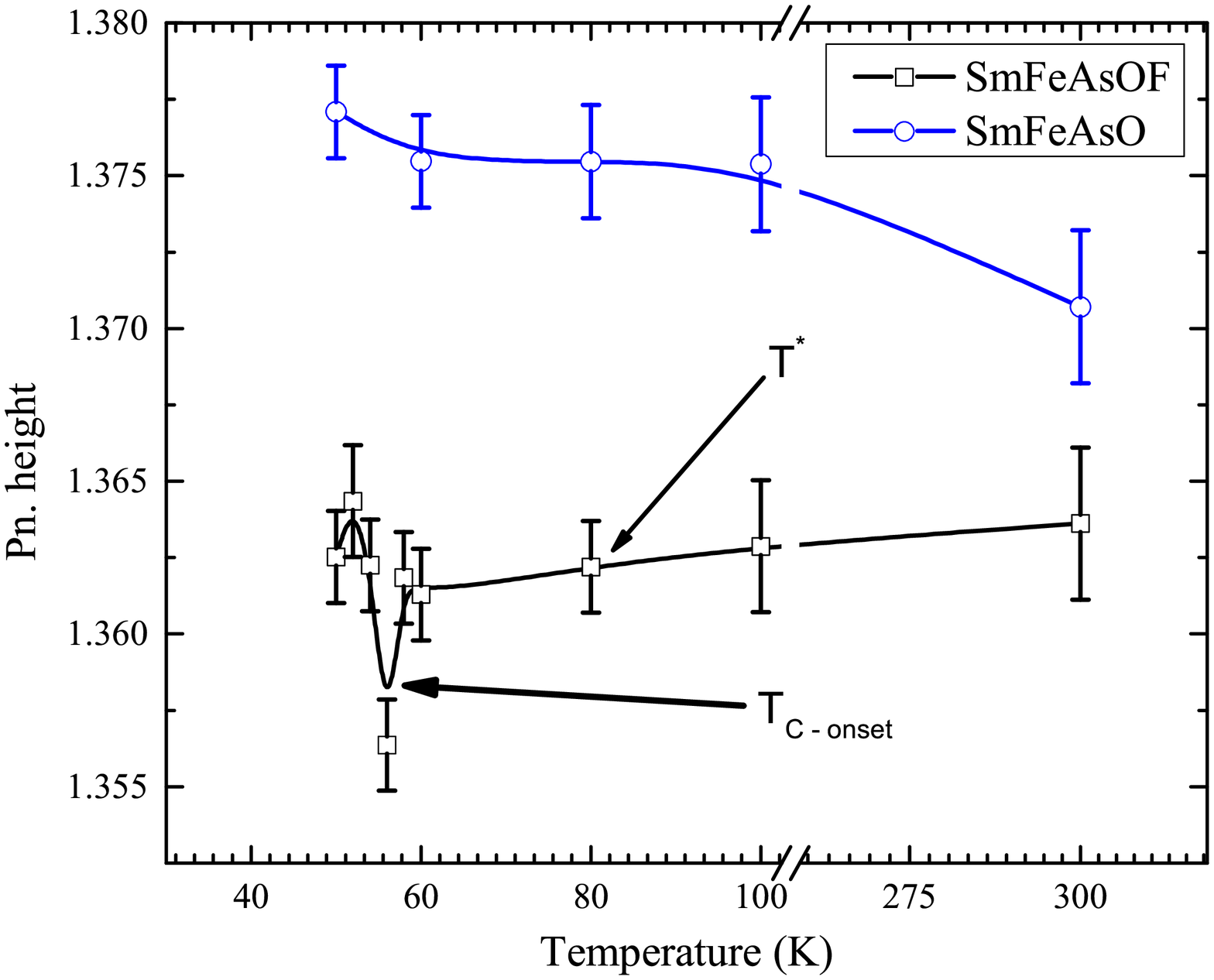}
\caption{\label{fig6}Temperature dependence of anion height $h$ in SmFeAsO and SmFeAsO$_{0.8}$F$_{0.2}$ samples.}
\end{figure}

Since $h$ was a parameter in the model used to fit EXAFS data, its variation as a function of temperature, in both, superconducting and non superconducting samples is shown in Figure \ref{fig6}. In the case of non superconducting sample, $h$ shows a slight increase in temperature from about 1.37 \AA~ at 300K to 1.377 \AA~ at 50K. Exactly opposite behavior is noted in the superconducting sample. Here, $h$, remains nearly constant down to about 80K and then starts decreasing with a minimum at 56K followed by a local maximum at about 52K. As in case of MSRD, the minimum in variation of $h$ corresponds with the onset of superconducting transition. This behavior of $h$ is akin to the behavior of the width of the Cu-O peak in the neutron radial distribution function (RDF) analysis in high T$_c$ superconductors \cite{mustre}. Such a variation has been interpreted to indicate presence of microscopic inhomogeneity below T$^*$.   The anomaly in the temperature dependence of $\sigma^2$ which begins at about 80K (Figure \ref{fig5}) also supports that electronic instability begins at this temperature. This clearly indicates the role of lattice distortions in superconductivity in Fe based superconductors.

\begin{figure}
\centering
\includegraphics[width=\columnwidth]{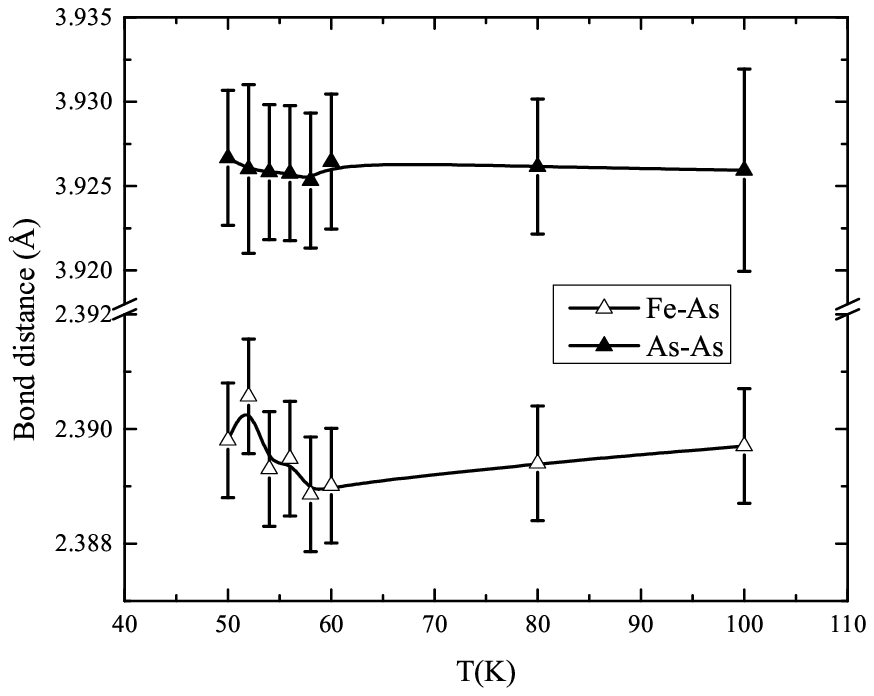}
\caption{\label{fig7} Thermal evolution of As-Fe and As-As bond distances in SmFeAsO$_{0.8}$F$_{0.2}$ superconductor.}
\end{figure}

Present results indicate an intimate connection in the mechanism of superconductivity in Fe based superconductors and that proposed in cuprates. It may also be mentioned here that a similarity between these two families of superconductors also exists in their phase diagrams, the close proximity of superconductivity to antiferromagnetic order and the presence of spin resonance peak in superconducting region indicating unconventional nature of superconductivity. In cuprates, EXAFS studies performed using polarized X-ray on bulk single crystal samples or single crystalline thin films have clearly shown a splitting ($\sim$ 0.1\AA) in Cu-O bond distances. Below T$^*$ the planar Cu-O bonds undergo dynamical elongation creating charge-rich and charge-poor domains. In order to observe the presence or absence of such domains in Fe based superconductors, more detailed experiments preferably on bulk single crystals are required.  However, we present the variation of nearest neighbor As-Fe and As-As bond distances in the superconducting compound in Figure \ref{fig7}. It can be seen that while the variation of As-As distance is nearly temperature independent, a very weak increase is noticed in Fe-As distance. It may be noted here that while As- As bond distance was a fitting parameter (lattice constant $a$), As-Fe bond distance was derived from $a$ and $h$. Therefore this increase is intimately related to change in pnictogen height as the sample approaches superconducting transition. Although, the increase in bond length is very small and within the estimated standard deviations, it is still significant due to different behaviors of $a$ and $h$ from which it is derived. This small increase suggests a close proximity in the mechanism responsible for superconductivity in these Fe based compounds and other families of superconductors like cuprates.

\section{Conclusions}
In summary, this paper reports temperature dependent As K edge EXAFS studies in superconducting SmFeAsO$_{0.8}$F$_{0.2}$ and non-superconducting SmFeAsO compounds. Local structural parameters like anion height, As-Fe bond distance and the mean square disorder in As-Fe bond distance of the fluorinated compound exhibit anomalous variation near the superconducting onset temperature. Such a variation in these parameters is absent in the non-superconducting compound. Furthermore, increase in Fe-As bond distance just below superconducting onset temperature indicates a similarity in distortions observed in high T$_C$ cuprates and these Fe based superconductors.

\ack{Authors thank Department of Science and Technology, Govt. of India for financial assistance under the project SR/S2/CMP-57. Thanks are also due to Photon Factory for beamtime under the proposal }

\end{document}